\documentclass{hotnets17}
\usepackage{times}
\usepackage{hyperref}
\usepackage{url}
\usepackage{graphicx}
\usepackage{xspace}
\usepackage{listings}
\usepackage{fancyvrb}
\usepackage{enumitem}
\setlist{nosep}

\hypersetup{pdfstartview=FitH,pdfpagelayout=SinglePage}

\newcommand{\eg}{e.g.,\xspace}
\newcommand{\ie}{i.e.,\xspace}
\newcommand{\etal}{et al.\@\xspace}

\newcommand{\parax}[1]{\noindent \textbf{#1:}}
\newcommand{\xref}[1]{\S~\ref{#1}}

\begin{document}

\title{An Internet Heartbeat}
\numberofauthors{2}
\author{
 \alignauthor Robert Beverly\\
 \affaddr{Naval Postgraduate School}\\
 \email{rbeverly@nps.edu}
 \and 
 \alignauthor Mark Allman\\
 \affaddr{ICSI}\\
 \email{mallman@icir.org}
}

\maketitle
\thispagestyle{empty}

\noindent \textbf{Abstract---} Obtaining sound inferences over
remote networks 
via active or 
passive measurements is difficult.  Active measurement campaigns face
challenges of load, coverage, and visibility.  Passive
measurements require a privileged vantage point.
Even networks under our own control too often remain
poorly understood and hard to diagnose.
As a step toward the democratization of
Internet measurement, we consider the inferential power possible were
the network to include a constant and predictable stream of dedicated
lightweight measurement traffic.  We posit an Internet ``heartbeat,'' which nodes
periodically send to random destinations, and show how aggregating
heartbeats facilitates introspection into parts of the
network that are today generally obtuse.
We explore the design space of an Internet heartbeat, potential
use cases, incentives, and paths to deployment.

\section{Introduction}
\label{sec:intro}

Much recent effort has gone into developing an understanding of the
presence and properties of \textit{all} end hosts on the
Internet using three primary strategies.  The first strategy is
active scanning, whereby probes are sent to every IPv4 address and
responses are collected \cite{heidemann2008census,
  durumeric2013zmap}.  The second strategy is for providers
with large footprints (e.g., Google or Akamai) to collect information
about the client devices (and their networks) that request service
\cite{Richter:2016:BCN:2987443.2987473, flach2016internet}.  
The final strategy
is to monitor ``background radiation'' arriving at
a swath of unused address space (network telescopes)
\cite{wustrow2010internet,
  dainotti11analysis, moore2004network}.  Each of these strategies is
helpful in understanding the Internet in-the-large, but each also
has significant drawbacks in terms of coverage, feasibility, visibility, and inferential
power.  For instance,
scanning probes are frequently considered abusive and blocked, are 
unlikely to traverse middleboxes, and require large amounts of
probing, while 
deriving an understanding from observing legitimate client requests
requires a massive global footprint, and hence the information is
concentrated in few hands.  Finally,
using telescopes depends on hosts accidentally sending
traffic to the unused address space being monitored.  

In this position paper we advocate for a new strategy: an
\textit{Internet Heartbeat} (IHB).  Rather than relying on
happenstance---e.g., a packet arrival at a darknet---or trying to
look in from the outside---e.g., trying to reach each host with a
probe---we explicitly bring all hosts into the measurement process by calling
for them to transmit a continuous, low-rate stream of messages into
the network.  This allows endpoints or observers along the path
to expect these messages and, hence, gain valuable insight into the
network.

At a high-level---which we refine below---any
node with a network stack can source and receive IHBs.  Participating hosts
periodically send a packet to a random destination that includes 
various metadata.  Observation
points---anything from a single host to a border router 
to a backbone link---can examine these packets to
facilitate introspection into myriad parts of the network,
including those that are today generally not observable and for which
there is no longitudinal data.  While
simple, we believe that an IHB is a powerful mechanism that
will enable operators, researchers,
and policy-makers to make more informed and complete
inferences over the network, as well as 
support new functionality.
Further, IHBs can directly benefit users who currently have only
limited insight into their network's reliability, performance, 
and availability, and still struggle to distinguish
between network problems originating locally, remotely, or with their
provider.
In particular IHBs enable:
\begin{itemize}
\itemsep 0pt
  
\item Any observer to have a global view akin to 
      large providers who make strong inferences from
      client
      traffic~\cite{Richter:2016:BCN:2987443.2987473, flach2016internet}..

\item
  Common measurements to be made passively---at the IHB
  receiver---without 
  issues of completeness, locality, or blocking.

\item  Continuity, enabling longitudinal
  measurement, inferences, and learning rather than a myriad of
  one-off surveys and on-demand measurement campaigns.
 
\item Stronger inferences, that are easier to reason over, than possible 
      via opportunistic measurements.

\item Users to gain insight into their own
  network's reliability, and performance, while
  providing diagnostic capabilities, thereby encouraging
  deployment.
\end{itemize}

In this position paper, we sketch the design space of an IHB and
consider its potential via candidate use cases, as well as 
overhead, feasibility, and deployment issues.


\section{Related Work}

Significant prior literature focuses on performing passive
measurement inferences.  In addition to legitimate traffic,
non-trivial levels of ``background radiation''
\cite{wustrow2010internet} arrive at networks due to self-propagating
malware, security scanners, and attacks.  Casado \etal
show the wealth of information that can be gleaned
passively~\cite{casado2005opportunistic}, while Durairajan \etal
leverage NTP server logs to estimate Internet latencies~\cite{durairajan2015time}.
Dainotti \etal
demonstrate how background radiation 
\cite{moore2004network} provides insight into global outage
and censorship events~\cite{dainotti11analysis}.  Finally, Sargent
\etal infer network policies from traffic arriving at darknets
\cite{SCAB15}.  While such
opportunistic measurement is powerful, analysis and inference is
complicated by the vagaries of attacks, the
spread and mitigation of malware, and what networks are affected.
In addition to generalizing opportunistic measurement, we
show that periodic IHBs permit stronger probabilistic inferences.

Individual networks frequently perform regular pair-wise
measurements between nodes or networks under their control, \eg 
Content Distribution Networks (CDNs) that run continual measurements to detect and route around
path problems~\cite{nygren2010akamai}.  
The IHB seeks to push such functionality deeper into the network stack
such that all networks are empowered with such knowledge without
having to implement their own application-layer protocols and
measurements.  As importantly, the IHB disseminates global knowledge
about the Internet, rather than focusing on an individual network.

Related in spirit to our heartbeats are BGP beacons \cite{mao2003bgp}.
BGP beacons are periodic announcements and withdrawals of specific
prefixes for the explicit purpose of measuring and understanding
real-world BGP behavior and dynamics.  BGP beacons have served as
enablers of important research efforts to understand and improve
routing, \eg \cite{Wang:2006:MSI:1159913.1159956}.  In a similar
fashion, our hope is that the IHB enables new measurements and
insights into the global Internet.

Finally, we note that many distributed systems employ the
``heartbeat'' notion---i.e., a
periodic signal to assert liveness~\cite{rfc4271,
bless-rfc5880, barborak1993consensus, vogels1996world} and achieve reliability.  A
key motivation for heartbeats is that a node cannot rely on
naturally occurring, event-driven protocol messages to ascertain
another node's health as silent failures or system changes cannot be
detected~\cite{kawazoe1997heartbeat}.  Our work extends the
heartbeat notion to the Internet.

\section{IHB Design}
\label{sec:design}

As early-stage work, we elect to explore various design alternatives.
In general, we follow designs that use
existing protocols, are
incrementally deployable (today), permit probabilistic inference, 
and minimize the security burden.

\subsection{Transport Protocol}
\label{sec:design:transport}

Choosing which transport protocol IHBs should use involves a complex
set of tradeoffs.  
A new transport protocol number would cleanly enable IHBs over IP, but
would likely cause IHBs to be discarded or treated differently by
today's network \cite{detal2013revealing}.  Similarly, adding IHBs as
an IP option is not feasible due to network
ossification~\cite{fonseca2005ip}.  While using UDP or TCP for IHBs
more closely mimics application traffic, it also brings increased
scrutiny from middleboxes and the possibility of unwanted
manipulation~\cite{detal2013revealing}.  
While ICMP is 
a well-established diagnostic and error reporting protocol, 
networks may treat ICMPs 
differently than
transports---like UDP or TCP---that applications often use. 

In our mock IHB implementation, we use ICMP with a 
new ICMP type, but plan to 
experiment with other transports.  
Our primary design criterion
is to design the IHBs such that they are 
readily and cheaply detectable, and therefore easy to block
or capture regardless of transport protocol. 



\ifdefined\blah
\subsection{Transport Protocol}
\label{sec:design:transport}

A new transport protocol number would cleanly enable IHBs over IP, 
but would likely cause IHBs to be discarded or treated differently by today's
network \cite{detal2013revealing}.
Instead, we consider three possibilities:

\parax{IP Option} 
This strategy allows for
flexibility in sending IHBs that mimic a variety of transport
protocols so that, unless a node was looking for the specific IHB IP
option, the packet would be handled per the policy for the given
transport.  Unfortunately, the ossification of the network has shown
that packets with any IP options are frequently blocked or, worse,
induce security alarms~\cite{fonseca2005ip}.

\parax{TCP} 
Unlike IP options, unknown TCP
options are generally ignored by the network and are not cause for
dropping traffic~\cite{honda2011still}.  This avenue would permit
traversing many
firewalls and middleboxes, however 
there are additional transports that we would like to use within the
IHB framework.

\parax{ICMP} A final option is to introduce a new ICMP type for
IHBs.  ICMP is well-established as a
diagnostic and error reporting protocol and, hence, is frequently
permitted through firewalls and not subject to deep intrusion
detection scrutiny.
However, as with inventing a new transport, ICMP is often treated
differently from normal user traffic and therefore the IHBs would
not directly share normal traffic policies.
\fi



\subsection{Meta-Information}


A heartbeat without any payload 
demonstrates
reachability and liveness from the source to the observation point.  However,
including additional meta-data aids
interpretation and
expands the possible inferences that can be drawn from IHB arrivals.
Here we describe four candidate pieces of IHB meta-data\footnote{IHBs
will be extensible to
accommodate new functionality.}.


\parax{Heart Rate} We include the rate
at which the host is sourcing IHBs and information about how
destination addresses are chosen (see \xref{sec:design:destination}).
While an observer may be able to
infer the sending rate of each IHB host, the period between heartbeats is
expected to be long enough---see next section---that this may be a
long and error-prone process, and especially brittle during failures
or outages.  Further, senders may wish to adapt the
rate at which they inject heartbeats dynamically, and on short time
scales.  By explicitly including the IHB
sending rate we can allow observers to set expectations
appropriately and stake inferences around the actual rate instead of
an estimated rate.

\parax{HostID} A well-known facet of the modern Internet is that IP
addresses serve as poor host identifiers---due to network address
translation (NAT), middleboxes, IPv6 privacy addresses, and aliases.
Therefore, we adopt the notion of a ``HostID'' from Allman \etal
\cite{allman2017principles} and include a small (\eg 16-bit) identifier in each
heartbeat payload.  The HostID is randomly generated by the source
of the IHB and can be changed over time\footnote{The meta-data could
include time since the HostID changed.}.  
A host identifier raises
immediate security and privacy concerns.  As observed in
\cite{allman2017principles}, these issues can be mitigated
by using a HostID that is purposefully small such that identifier
collisions are common.  This means the HostID is meaningless in a global sense,
but must be coupled with the IP address to draw inferences about a
host's behavior.  For instance, the HostID permits an approximation
of the number of hosts behind a NAT.

\parax{Originating TTL} Including the TTL with which the heartbeat
was sent removes ambiguity of scope and allows both the recipient
and observers along the path to determine how far the heartbeat has
traveled.  We utilize TTL values to infer path changes in
\S\ref{sec:cases:paths}.

\parax{Timestamp} The timestamp permits one-way latency approximation
when the source and observer's clocks are synchronized, but further
allows observers to note latency \emph{changes} over time even without
synchronization.

\subsection{Heartbeat Destination}
\label{sec:design:destination}


The intuitive notion is that IHBs should be transmitted around the
network uniformly at random.  We consider two aspects of choosing the
destination: ($i$) the pool of addresses from which to choose; and
($ii$) the order of using the addresses in the pool.  Here we
consider three pools:

\parax{Entire Address Space} This pool consists of all $2^{32}$ IPv4
addresses.  In this case, the goal is for all hosts to exchange IHBs
with all other hosts on the network.  Assuming one host per IP
address and full address usage, this goal requires $2^{64}$ IHBs.
Note, however, that the number of IHBs required for
full coverage decreases in proportion to the size of the observation
point lens, e.g., as discussed in detail in \S\ref{sec:overhead}, 
a large darknet (monitoring /8
of IPv4 address space) can expect to 
receive a heartbeat from an arbitrary host after every
$\frac{2^8}{2} = 128$ heartbeats it sends.

\parax{All /24 Networks} This pool consists of all $2^{24}$ IPv4 /24
network blocks.  This strategy leverages the fact that a /24 address
block is the smallest block that can be confidently routed across
the Internet.  Therefore, IHBs to multiple hosts within a /24
largely share fate.  While this is not always true---\eg end host
firewall configurations can vary across hosts within a /24---the
relaxed goal results in $2^{48}$ IHBs to satisfy the goal---or,
$\frac{1}{64K}$-th of the number needed for all hosts to exchange
IHBs.  Another benefit of using a /24 pool is that all IHB hosts
within a /24 could \textit{cooperate} to meet the goal.  For instance, if
32 hosts within a /24 send IHBs then each would have to send 524K
($\frac{2^{24}}{2^{19}}$) messages to cover the remote /24 address blocks.

\parax{Local Subnet} This pool consists of the host's
local network, and would be used as a \textit{Local Heartbeat}
(LHB).  For instance, a typical home network using the private 
192.168.0.0/24 subnet would have 255 different LHB target addresses,
while the home network's public gateway (connected to the provider)
might be part of a /22 subnet and, hence, have 1024 LHB targets.  In
general, LHBs would be sent with a restrictive TTL and be used for local 
debugging and diagnostics.  A refinement of the LHB idea is to make 
the TTL inversely proportional number of matching most significant
bits between the source and the randomly chosen target.

\noindent Next, we consider the order in which hosts are probed:

\parax{Pure Random} Choosing an entirely random destination from the pool
is simple to implement, ensures uniform coverage in time and space,
and facilitates inferences with confidence bounds.\footnote{With
  /24 pools, a /24 is chosen at random and then we
  independently choose the low order octet of the address at
  random.}

\parax{Random Permutation} Rather than choosing addresses at random, the
sender could instead randomly permute the entire pool.  The random
permutation ensures that every destination is used before any
destination is repeated.  Additionally, the sequence can
deterministically repeat once the host has sent to every address,
thereby adding additional predictability to the potential
inferences.  

The above discussion is in terms of IPv4 address space.  
Sending IHBs over IPv6 networks will require
understanding the general usage of address blocks (as discussed in
\cite{CLAB16}), understanding which blocks of IPv6 addresses are
routed or having some form of carefully curated hit-list.  We
discuss possibilities for IPv6 heartbeats in~\S\ref{sec:ipv6}.

\subsection{Source Address}
\label{sec:design:source}

Many Internet-connected devices have multiple interfaces, e.g., 
WiFi and cellular, while routers have more than one interface by
definition.  We explicitly envision routers participating in the IHB
via their control plane network stack.  To expose these interfaces,
and the paths they utilize, the IHB should effectively run independent
instances of the heartbeat protocol on each physical or virtual
interface, and use the interface's assigned IP address as the heartbeat
source.  However, the IHB meta-data, in particular the HostID, should
be consistent among interfaces.  

\subsection{Rate}

The rate at which a device sources IHBs is a local policy decision
and is explicitly included within the heartbeat meta-data so that
observers can form concrete expectations over when the next
heartbeat should arrive.  A possible enhancement to this strategy is
to source two back-to-back heartbeats for each destination.  While
doing so doubles the data rate---or halves the number of IHB
recipients per time unit---it enables basic packet-pair dispersion
techniques for \eg estimating path capacity
\cite{dovrolis2001packet}.

\subsection{Integrity}

As a connectionless protocol, the IHB as
specified is trivially spoofed.  An adversary could source heartbeats
with a spoofed source IP address in order to pollute inferences, for
instance to make it appear as though a network is up when it is, in
fact, suffering an outage.  While ingress filtering
\cite{bless-rfc2827} can
mitigate the ability to spoof in some cases, its deployment is
not ubiquitous and spoofing remains a present 
concern~\cite{Beverly:2009:UED:1644893.1644936}. 
However, an observation point that receives conflicting heartbeat
information (arrival rate, TTL, or HostID) from a given source or network can infer that some of the
IHBs 
are illegitimate (and should be ignored).  

Even stronger forms of integrity protection may be feasible.  For
instance, while we do not wish to depend on public key cryptography
(for reasons of speed and PKI deployment obstacles), we could utilize
heartbeats themselves to distribute shared secrets.  Here, heartbeats
would include a per-target or per-/24 random key in the meta-data.
When subsequently sourcing heartbeats, the host would include a keyed
integrity check (such as an HMAC) using the previously received key.  
In this fashion,
an adversary can still spoof heartbeats, but cannot do so undetected
unless she has access to the heartbeats arriving at the remote
network.  
Alternatively, IHBs may include \emph{chained} integrity
using ephemeral secrets, where the integrity key to the $i$'th IHB
is contained in the $i+1$'th heartbeat.
We plan to explore options for heartbeat integrity more
completely in future work.

\section{Use Cases}
\label{sec:cases}

To shed light on potential inferences possible with an IHB,
we explore three use-cases: outage detection, residential broadband 
debugging, and 
forwarding path changes.

\begin{figure}[!t]
  \centering
  \resizebox{1.0\columnwidth}{!}{\includegraphics{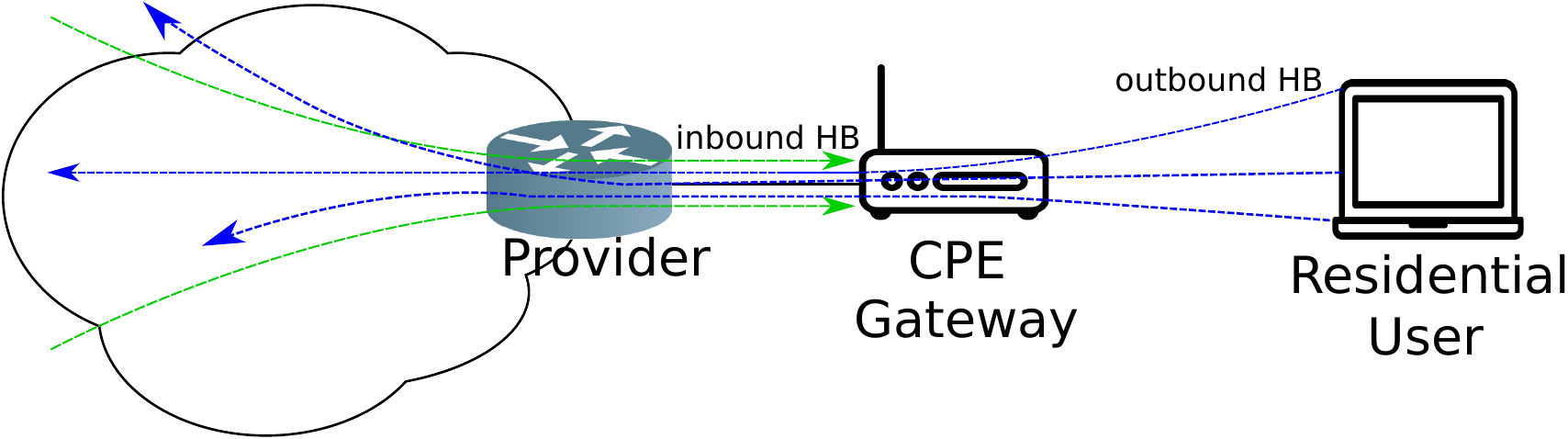}}
  \vspace{-7mm}
  \caption{Inbound and outbound IHBs enable users to better
  diagnose problems and localize their origin, while facilitating
  longitudinal measurements of their provider's availability
  and reliability.}
  \label{fig:residential}
  \vspace{-4mm}
\end{figure}

\subsection{Outage Detection}
\label{sec:cases:outage}

An increasingly important global
measurement task is identifying Internet outages.  Traditional methods
of outage detection rely on passively monitoring BGP
\cite{Dhamdhere:2007:NTN:1364654.1364677} or performing large-scale
active measurements \cite{quan13trinocular}.  Unfortunately, many
outage events do not present as BGP activity, while active
measurements require significant and continual probing.  In contrast, an IHB
inherently provides a means for understanding the data-plane
reachability of remote networks simply by making inferences over the
sequence of heartbeats received.  

Naturally, different vantage points will be able to make stronger, and
faster, inferences than others.  For example, consider again a /8
darknet where we wish to monitor the reachability of a remote /24
subnet.  Assume that only a single host on the remote /24 participates
in the IHB at a rate
of 1~pps.
Thus,
every second there is a
$\frac{1}{2^8}$ probability that the node sends a heartbeat to a
destination in the /8 monitored by the darknet. 
After $k$ seconds 
the the probability that the /8 receives no heartbeats from this host
is: $\left( 1 - \frac{1}{2^8}\right)^k$.  Thus, the probability that
the remote /24 is reachable even though the darknet has received no
heartbeats after $k=100$
seconds
is approximately 67\%.
However, the likelihood of not receiving a heartbeat by chance rapidly
decreases over time: there is only a $3\%$ chance that the /24 is reachable
after receiving no heartbeats for
15 minutes.

Of course, more hosts participating in the IHB on the remote /24
increase the outage sensitivity and confidence, while smaller darknets
decrease detection ability.  The rough numbers in the preceding
example are meant to highlight what is possible given realistic
assumptions over IHB adoption and the size of today's darknets.  The key
point is that the periodic behavior of IHBs permits reliable
probabilistic inference where conscious engineering tradeoffs can be
made given the required detection sensitivity.


\subsection{Residential Broadband}
\label{sec:cases:residential}

We want IHBs to benefit not just
researchers and operators, but also the 
hosts and---most importantly---users who participate in the
sourcing and receiving of heartbeats.  As an example of aligning
IHBs with user incentives, we consider IHB use within residential networks. 

While there has been significant recent effort toward understanding
residential broadband speed and
performance~\cite{bauer2010understanding,kreibich2010netalyzr}, 
less is known about the
reliability of these networks.  This is particularly true for
broadband customers who: ($i$) will only notice network failures if they
occur when the user is home and trying to use the network; ($ii$) do not
maintain long-term statistics over their connection's reliability or
provider's availability; and ($iii$)) cannot distinguish between a network
failure within their home, in the access network, or occurring remotely
(i.e., outside of the control of the provider).  IHBs provide a means
to enable continuous monitoring of a customer's home access network,
while introspection of IHBs facilitate diagnostics and the ability to
measure adherence to service level agreements.

Assume that a broadband user's home CPE gateway inspects both inbound
and outbound heartbeats and further computes statistics and metrics
over the heartbeats for the user.  In Figure~\ref{fig:residential},
inbound IHBs and LHBs arrive at the customer's CPE (in
\S\ref{sec:overhead}, we
explore the expected rate of incoming heartbeats---the salient
feature here is that a constant stream of heartbeats arrive at the
CPE).  Further, the CPE observes outbound heartbeats sent by host(s)
on the user's local network (either IHBs or LHBs).  These outbound
heartbeats provide evidence that the user's clients are able to obtain
an address and reach the gateway.  

The heartbeats immediately provide valuable information to the CPE,
which can be exposed to the user (e.g., through a user-friendly web
interface).  Employing the same probabilistic methods as for outage
detection (\S\ref{sec:cases:outage}), the CPE can thus aid in
isolating whether reachability issues are local, on the user's access
connection, within the user's provider, or remote.   We further
envision the CPE providing information to the user over: ($i$) the
reliability of the connection to her provider; ($ii$) evidence of which
networks could reach the user at a given point in time; and ($iii$)
evolution of latencies from particular hosts and networks.  

These inferences can be further strengthened by obtaining additional
evidence from other hosts on the same subnetwork through LHBs if these
LHBs work to distribute knowledge about IHBs received among all hosts
on the subnet.  

\subsection{Path changes} 
\label{sec:cases:paths}


Heartbeats can reveal both global and specific forwarding dynamics
within the network via simple analysis of the IP time-to-live (TTL).
Because routers decrement the TTL before forwarding the packet, the
TTL provides an approximation of the number of hops from the source to
the observation point.  TTL changes for the same source or network are
thus indicative of routing changes or load-balancing, allowing
heartbeats to serve as a path change detection mechanism.
Importantly, aggregation and correlation of heartbeats can identify
networks that share paths and share fate.  

As an expository example, we sent heartbeats from 26 Planetlab
nodes~\cite{chun2003planetlab}
toward random addresses of a darknet over a six day period.  While the
TTL of received heartbeats from 16 of the nodes remained constant over the
collection period, the TTL changed for 10 of the sources.  Heartbeats from three of
the ten planetlab sources had an equal distribution of TTL values,
revealing the presence of load-balanced paths.  A time-correlated
heartbeat TTL change was observed among three of the sources and
lasted for approximately one hour.  Because this path change did not
affect all heartbeats, the routing change was distant to the telescope.
However, the correlated change suggests that the three nodes share a
path in the network core, and illustrates the potential inferential
power of tomography techniques \cite{Dhamdhere:2007:NTN:1364654.1364677}
combined with IHBs.

Last, path changes can serve as a rough indicator of potentially
malicious traffic.  While IP source address spoofing
\cite{Beverly:2009:UED:1644893.1644936} allows an attacker to
impersonate an address, the attacker cannot set her packet's initial
TTL such that she is topologically closer than the number of router
hops.
As such, packets that arrive with a TTL less than
previously observed from that network may be due to a routing change, or may indicate
spoofing.  If heartbeat packets arrive from a host with one TTL while
TCP SYN packets or DNS packets arrive with a different TTL, this
further suggests the presence of spoofing.  

\subsection{Additional applications}


The preceding use cases highlight possible
applications of IHBs; we believe they can enable other uses including:

\parax{Traffic optimization} A well-known property of Internet
        routing is that alternative paths and indirection can
        provide superior end-to-end
        performance~\cite{Andersen:2002:RON:510726.510740}.  CDNs and service
        providers therefore routinely perform measurements to 
        inform their path selection \cite{nygren2010akamai}.  IHBs expose
        information (loss, capacity) to receivers about available
        paths and their performance. 
        We envision network overlays (\eg DHTs, peer-to-peer, and Tor)
        using IHBs to inform their overlay construction without the
        need for dedicated active probing. 

\parax{Census}  Network census campaigns, \eg
\cite{heidemann2008census}, have seen significant 
        interest in applications ranging from vulnerability analysis
        \cite{durumeric2013zmap} to hitlist generation
        \cite{fan2010selecting} to estimating
        address space usage \cite{dainotti2013estimating}.  Whereas these existing techniques
        require significant active probing (where load is concentrated
        at the probing host) and suffer from
        completeness and coverage issues, an IHB naturally facilitates
        these studies in a highly distributed and continual fashion.

\parax{Alias resolution} In \S\ref{sec:design:source}, we note 
        that devices with
        multiple interfaces should run an IHB instance on each.
        Because the HostID is tied to the host rather than the
        interface, we imagine the ability to perform probabilistic
        alias resolution, the process of identifying the set of IP
        addresses belonging to a single physical device such as a
        router~\cite{keys2013internet}.  Whereas alias resolution is
        today an expensive, time-consuming, incomplete, and
        error-prone task~\cite{keys2013internet}, a passive
        heartbeat observation point can easily identify candidate
        aliases by clustering source addresses of heartbeats with
        the same HostID.  Naturally, this set of candidate aliases
        will initially be very large---due to the small HostID
        identifier space---but can be refined over time as stacks
        choose new random HostIDs or can be combined with other
        information such as traceroute data to detect aliases.

\parax{Policy inference}  Work on understanding
        differential traffic treatment or blocking frequently uses
        multiple vantage points to compare results
        \cite{Burnett:2015:ELM:2785956.2787485}.
        Similarly, comparing IHBs that arrive at different
        collection points can shed light into different network
        policies\footnote{Here, the choice of transport protocol
        (\S\ref{sec:design:transport}) is
        important so that heartbeats are classified in the
        same way as normal traffic.}.  For instance, if we observe IHB packets
        arriving from source network $S$ at destination network $X$,
        but no IHB packets arriving from $S$ to network $Y$, then
        something is blocking traffic from $S$ to $Y$.  Using
        additional IHBs and network tomography 
        \cite{Dhamdhere:2007:NTN:1364654.1364677} 
        may further isolate network policy.

\section{Discussion}

Although the IHB is largely a thought experiment at this time, we
discuss
practical deployment issues next.

\subsection{Overhead}
\label{sec:overhead}

To demonstrate the back-of-the-envelope feasibility of IHBs with
respect to overhead, 
we consider the magnitude
of IHB traffic using reasonable assumptions over the expected number
of participants, data rate, and receiver size.  

Let the IHB observation point monitor a prefix with mask $m$ (\ie 
for a single IPv4 host $m=32$).
The probability of a single IHB participant sending a heartbeat that arrives
at this observation point is then: $p = \frac{1}{2^{m}}$.

Assume that there are $n$ hosts participating in IHB and 
sending IHB packets.  For simplicity, assume that each
host sends a single IHB in each epoch, \ie they all send
at the same rate.  Then,
the expected number of IHB packets arriving at the observation
point in a single epoch is:
$\frac{n}{2^{m}}$.  

Next, assume that each source
sends heartbeats to random destinations at a rate of $r$ packets
per second.  Then, the expected arrival rate of IHB packets is: 
$a = \frac{nr}{2^{m}}$.  

Assume that one-quarter of the IPv4
Internet addresses participate in IHB, \ie
$n=2^{32-2}$.\footnote{With NATs this is $\ll0.25$ 
  of devices connecting to the Internet.}
If we further assume a relatively low per-host IHB rate of
$r=\frac{1}{8}pps$, we can estimate the expected traffic load on
the observation point due to heartbeat traffic.  For instance, at
the two extremes of a single end host versus a network telescope:
\begin{itemize}
 \item End-host ($m=32$): $a = \frac{2^{-3}(2^{30})}{2^{32}} =
       \frac{1}{32}pps$
 \item Telescope ($m=8$): $a = \frac{2^{-3}(2^{30})}{2^{8}} = 
       2^{19} \simeq 500kpps$
\end{itemize}

Thus, given relatively conservative assumptions over IHB deployment,
we observe that a single host receives a negligible rate of heartbeat
traffic, while a large aggregation point such as a /8 telescope
receives a technically realistic and reasonable half-million IHB 
packets per second.

\subsection{Security and Privacy}

A consequence of increasing the visibility of hosts on the
network is the potential for this information to be used in
unscrupulous ways.  For instance, a heartbeat provides an
explicit indication of a remote node's liveness at a given instant in
time, thereby providing attackers a potential target.  We argue,
however, that heartbeats do not enable a new attack vector, as
adversaries are already capable of high-rate,
exhaustive vulnerability scanning \cite{durumeric2013zmap} and make extensive use
of available hitlists \cite{fan2010selecting, shodan}.  In practice, protecting
end hosts from attacks is an issue orthogonal to liveness.
Nonetheless, networks that wish to remain outwardly ``dark'' 
can easily maintain this particular security
posture as IHB packets are designed to be easily
identified so that their transmission and reception can be
blocked as required.  


Finally, heartbeats raise potential concerns over tracking and
privacy.  We first note that heartbeats are designed to illuminate the
network, and there is no identifier that persists across networks (as
might be required to facilitate tracking).  And because heartbeats are
periodic, they provide no explicit indication of user or host
activity.  While it may be possible to discern when a particular host
is powered on or off, this information is again readily available via other
methods, e.g., \cite{schulman2011pingin}.  Indeed, web cookies, software
updates, and other chatty protocols leak significantly more private
information than IHB \cite{acar2014web}.  As such, we believe the additional 
security burden imposed by heartbeats to be minimal and
manageable, especially in relation to their potential benefit.

\subsection{Deployment and Incentives}

IHBs help support global measurement and promote a better
understanding of the network, and we sketched several 
motivating examples of how they might be leveraged.  Here we 
envision possible incentives and paths to deployment.

\parax{DNS servers} As an initial step toward global deployment, we
target DNS servers as ideal candidates to source IHBs for several
reasons.  First, DNS roots, authorities, and public resolvers are
well-known infrastructure that is globally distributed, continually
alive, and well-connected.  Second, users depend on these servers as
critical infrastructure.  Thus, both users and providers have a vested
interest in understanding, optimizing, and debugging the availability
and performance of DNS connectivity.

\parax{End-users} Second, we believe IHBs can provide direct benefit
to participating users as described in \S\ref{sec:cases:residential}.
While the benefit increases in proportion to the number of
participants for IHBs, deployment can be bootstrapped with DNS
servers, while end-users can benefit from LHBs immediately as
deployment progresses.  
We envision IHB support within home routers
and CPE to be critical for widespread adoption and plan to 
implement a prototype that can run on OpenWRT as a first step toward
understanding how users can benefit.

\subsection{IPv6}
\label{sec:ipv6}


Heartbeats may be especially important in IPv6 where the size of the
128-bit address space poses unique measurement obstacles.  For
instance, whereas exhaustive vulnerability, census, and topology
scanning,
\eg \cite{durumeric2013zmap,
imc16yarrp}, 
are infeasible in IPv6, 
an IPv6 IHB can naturally illuminate active hosts,
networks, and paths.  However, the sparsity of the IPv6 address space implies
that choosing the heartbeat target (destination IPv6 address) 
at random will rarely produce an
active host, much less a routed network.

With only 0.002\% of the IPv6 address space 
advertised currently \cite{potaroo}, only approximately one in every 50K
random heartbeats will be directed toward an advertised IPv6 prefix.
While randomly destined IPv6
heartbeats will be observed at darknets,
the probability of reaching
an active IPv6 host is vanishingly small (as small as $O(2^{-64})$ in the case of
privacy preserving addresses~\cite{bless-rfc4941}).  
We therefore 
advocate for a
reserved /64 destination address
suffix for IHB.  In this way, network borders and telescopes can readily
identify IPv6 IHB traffic, or special end-network routing may be
installed for IHBs.  Further, it may be possible to develop 
mechanisms that allow all IPv6
IHB nodes on a network segment to receive the heartbeats, for instance
via a special MAC address mapping or rewriting IHB packets into
link-local IPv6 multicast.  

Further enhancements may be possible to bias IPv6 IHBs to active 
and routed portions of the address space.  For instance,
we observe that received heartbeats can be used to
discover and learn the active address space.  For example, the /64
network corresponding to the source address of received heartbeats
is certain to be routed, and clustering of contiguous /64s can produce 
larger prefixes.  
IPv6 hosts
participating in the IHB can bias their probing to these learned
blocks to better utilize the probing budget.  While this learning 
process might seem to require significant time (and wasted probing),
we note that the control plane of routers participating in IHB have 
direct and accurate knowledge of the routed IPv6 address space, and
can effectively act as seeds to bootstrap the learning process.

\section{Future Work}

\ifdefined\rehash
Today's Internet is a distributed system that suffers from a lack of
visibility---nominally because of its sheer size.  A node or network
on the Internet may have a notion of the state of remote networks with
which it is actively communicating, but has little to no insight into
the rest of the network.  This lack of visibility is underscored by
the need to perform significant active scanning to gain even an
approximate census of the nodes on the network
\cite{heidemann2008census, durumeric2013zmap}, or the need for CDNs
\cite{nygren2010akamai} to perform server selection on behalf of blind
clients.  This work proposes Internet Heartbeats as a new mechanism
that can complement existing network measurement techniques, and allow
wider introspection and understanding of the network.
\fi

Significant future work remains to
realizing the IHB vision, and initial architectural choices may prove
important.  
%
For instance, the current design 
requires a non-trivial number of IHBs before every node has
touched every other node.  Instead, a possible alternative is to 
coordinate heartbeats among participating nodes on a network
segment---e.g., via a local gossip mechanism.  In this way,
nodes could divvy up the random permutation to minimize coverage time.

Unlike more radical network architecture proposals, implementing and
experimenting with IHBs is readily possible, and we have begun just
this.  While IHBs could be placed into the network stack of operating
systems, our approach is to develop a user-space program,
\texttt{heartbeatd}, that implements IHB.  We plan a small-scale
deployment to perform
long-term continual heartbeats.  Simultaneously, we plan to explore
the real-world inferences possible at a variety of observation points,
including at individual nodes, a network border, CPE, and on a darknet.



\clearpage
\newpage
\small{
\bibliographystyle{abbrv}
\bibliography{heartbeat,mallman,rfc}
}

\label{lastpage}

\end{document}